\documentclass{iau}  
\usepackage{graphicx}
\usepackage{natbib}
\usepackage{hyperref}

\title[Machine-learning investigation of dusty evolved stars with metallicity] %% give here short title %%
{Using machine learning to investigate the populations of dusty evolved stars in various metallicities}

\author[Maravelias \textit{et al.}]   %% give here short author list %%
{Grigoris Maravelias$^{1,2}$\thanks{contact: maravelias@noa.gr}, 
Alceste Z. Bonanos$^1$, 
Frank Tramper$^3$, 
Stephan de Wit$^{1,4}$, 
Ming Yang$^{1,5}$, 
Paolo Bonfini$^6$, 
Emmanuel Zapartas$^1$, 
Konstantinos Antoniadis$^{1,4}$, 
Evangelia Christodoulou$^{1,4}$
\and Gonzalo Muñoz-Sanchez$^{1,4}$
}

\affiliation{
$^1$IAASARS, National Observatory of Athens, Greece % GR-15236, Penteli, Greece    
\\[\affilskip]
$^2$Institute of Astrophysics, FORTH, Greece %GR-71110, Heraklion, Greece
\\[\affilskip]
$^3$Institute of Astronomy, KU Leuven, Belgium %Celestijnenlaan 200D, 3001, Leuven, Belgium
\\[\affilskip]
$^4$Department of Physics, National and Kapodistrian University of Athens, Greece %Zografou, Greece, 
\\[\affilskip]
$^5$Key Laboratory of Space Astronomy and Technology, National Astronomical Observatories, Chinese Academy of Sciences, People’s Republic of China 
\\[\affilskip]
$^6$Ballista Technology Group, Florida, USA
}

\pubyear{2022}
\volume{361}  %% insert here IAU Symposium No.
\jname{Massive Stars Near and Far}
\editors{N.~St-Louis, J.~S.~Vink \& J.~Mackey, eds.}
\begin{document}

\maketitle

\begin{abstract}
Mass loss is a key property to understand stellar evolution and in particular for low-metallicity environments. Our knowledge has improved dramatically over the last decades both for single and binary evolutionary models. However, episodic mass loss although definitely present observationally, is not included in the models, while its role is currently undetermined. A major hindrance is the lack of large enough samples of classified stars. We attempted to address this by applying an ensemble machine-learning approach using color indices (from IR/\textit{Spitzer} and optical/Pan-STARRS photometry) as features and combining the probabilities from three different algorithms. We trained on M31 and M33 sources with known spectral classification, which we grouped into Blue/Yellow/Red/B[e] Supergiants, Luminous Blue Variables, classical Wolf-Rayet and background galaxies/AGNs. We then applied the classifier to about one million \textit{Spitzer} point sources from 25 nearby galaxies, spanning a range of metallicites ($1/15 - \sim3~Z_{\odot}$). Equipped with spectral classifications we investigated the occurrence of these populations with metallicity.

%% add here a maximum of 10 keywords, to be taken form the file <Keywords.txt>
\end{abstract}

\firstsection % if your document starts with a section,
              % remove some space above using this command.

%%%%%%%%%%%%%%%%%%%%%%%%%%%%%%%%%%%%%%%%%%%%%%%%%%%%%%%%%%
\section{Introduction}

Although our knowledge on the modeling of stellar evolution for massive stars has  significantly advanced over the last couple of decades, we still struggle to understand the influence of some key parameters. One such example is mass-loss, for which the theoretical predictions are at odds with the values derived from observations. This becomes more evident as we move away from the main-sequence \citep[e.g.][]{Smith2014}. Certain phases of evolved massive stars display - additionally to the stellar winds \citep[e.g.][]{Vink2021} - episodic mass-loss activity, such as in Luminous Blue Variables (LBVs; e.g. \citealt{Davidson2020}), Wolf-Rayet stars (WRs; e.g.  \citealt{Crowther2011}), B[e] supergiants (BeBRs; e.g. \citealt{Kraus2019a}), Yellow Supergiants (YSGs; e.g. \citealt{deJager1998}), Red Supergiants (RSGs; e.g. \citealt{Levesque2017}). We already know that the majority of the massive stars are born into multiple systems \citep[e.g.][]{Kobulnicky2014, Sana2014} which results in a significant fraction of binary interactions \citep[e.g.][]{deMink2014}. However, binarity may not be the only channel for episodic mass-loss. Moreover, it is only now that we start to explore these populations in low metallicity environments, which can help us understand better the underlying mechanisms \citep[e.g.][]{Lorenzo2022}. Simultaneously, it is those environments that seem to host superluminous supernovae, which may be the result of the interaction of ejecta with a complex dusty circumstellar environment formed by possible multiple mass-loss events prior to the explosion \citep{Neill2011,Gal-Yam2012}.
 
The ASSESS project\footnote{\url{http://assess.astro.noa.gr/}} (P.I. Alceste Bonanos, see also \citealt{Bonanos2022} this volume) aims to understand the importance of episodic mass-loss by studying a large number of evolved stars in nearby galaxies. One of the approaches is to use machine-learning methods to classify sources without any spectral classification, and to perform follow-up observations to verify their nature. Given this tool we can also perform a study of those populations across different metallicity environments. The current paper summarizes our first results towards this goal.

\section{The classifier}

Since the classifier is presented in detail in \cite{Maravelias2022}, we provide only a brief overview in this section, so that the reader understands the post-processing steps and the final results. 

After careful search in literature we compiled a \textit{Gaia} cleaned sample of 932 M31 and M33 sources with known spectral types. We grouped these into seven classes of interest, i.e. in BSG, YSG, RSG, BeBR, LBV, (classic) WR, and background galaxies (GAL; as a collective class of outliers). We used photometry, and in particular consecutive color indices, from the \textit{Spitzer} ([3.6] and [4.5]) and Pan-STARRS ($r, i, z, $  and $y$)  bands (not corrected for extinction as there is not enough data to perform this and the influence in those selected bands is minimal). As this initial sample is highly imbalanced we employed synthetic data approaches to overcome the limitations imposed by the underrepresented classes. Although this does not completely alleviate the problem with LBV it helps with other classes (e.g. WR). Our classifier is an ensemble approach of three supervised machine-learning algorithms, that is Support Vector Classification, Random Forests, and Multilayer-Perceptron (a type of neural network). Their output results probabilities per class and these are combined in order to obtain the final probability distribution, while the final class is derived by using the class that corresponds to the maximum probability. 

The results from the application of this approach are fairly good. We managed to correctly identified from the initial training sample $\sim60\%$ to $\sim80\%$ for BSG, GAL, and YSG, $\sim73-80\%$ for BeBR, and  $\sim45\%$ for WR, with the best results obtained for the RSG ($\sim94\%$) and the worst for LBV ($\sim28\%$ - only from the Support Vector Classification). These results are comparable or slightly better when compared to the results from \cite{Dorn-Wallenstein2021}, who worked with a much less homogeneous (with respect to the labels) but more populated Galactic sample. Our overall weighted balanced accuracy is $\sim83\%$. 

Our test sample consisted of a (limited) number of sources collected for IC 1613, WLM, and Sextans A galaxies. Due to the fact that $\sim14\%$ of these sources have missing values (in the selected feature space, i.e. not having measurements across all selected \textit{Spitzer} and Pan-STARRS bands), we performed data imputation by replacing the features' values using iterative imputer (which models, iteratively, each feature with missing values as a function of other features, and uses that estimate for imputation). Although overall the missing data imputation does not affect much the results on this particular test dataset, our results showed that we can retain a relative good prediction score even when missing up to three features (out of the total five available, i.e. the consecutive color indices). The accuracy obtained for this test sample (at $\sim70\%$) was lower than that for the M31 and M33 sample. We attribute this discrepancy partly to photometric issues and to the total effect of metallicity, which might affect the intrinsic colors of the sources and extinction due to the different galactic environments. However, we still lack a statistically significant sample to study and understand the influence of the metallicity, a target towards which the ASSESS project is actively working to address, both through the machine-learning applications such this classifier as well as dedicated observational campaigns (see \citealt{Munoz-Sanchez2022, deWit2022} in this volume).

\section{Application and Results}

After completing the training of the model using the whole available sample of sources in M31 and M33, we proceeded with its application. We used source lists for 25 nearby  galaxies, spanning a range of 1/15 to $\sim3~Z_{\odot}$, that have been compiled from \textit{Spitzer} photometric catalogs, and after \textit{Gaia} screening. The classifier returns spectral types for 1156925 sources in total, including information on band completeness (the fraction of color indices without missing data) and probability. From all these sources only a tiny $\sim0.5\%$ have spectroscopic classifications.

We applied a number of selection cuts in order to first examine secure candidate sources. For this we employed a limit for band completeness so that we keep sources with measurements across all bands (i.e. band completeness equal to 1), which means that we excluded all sources with missing values. Then we considered only sources with a classification probability of at least 0.51. This value corresponds to the peak of the incorrect classifications for the M31 and M33 sources \citep[see Fig. 6][]{Maravelias2022}. 

\begin{figure*}
\begin{center}
 \includegraphics[trim= 0 80 0 130,clip, width=\textwidth]{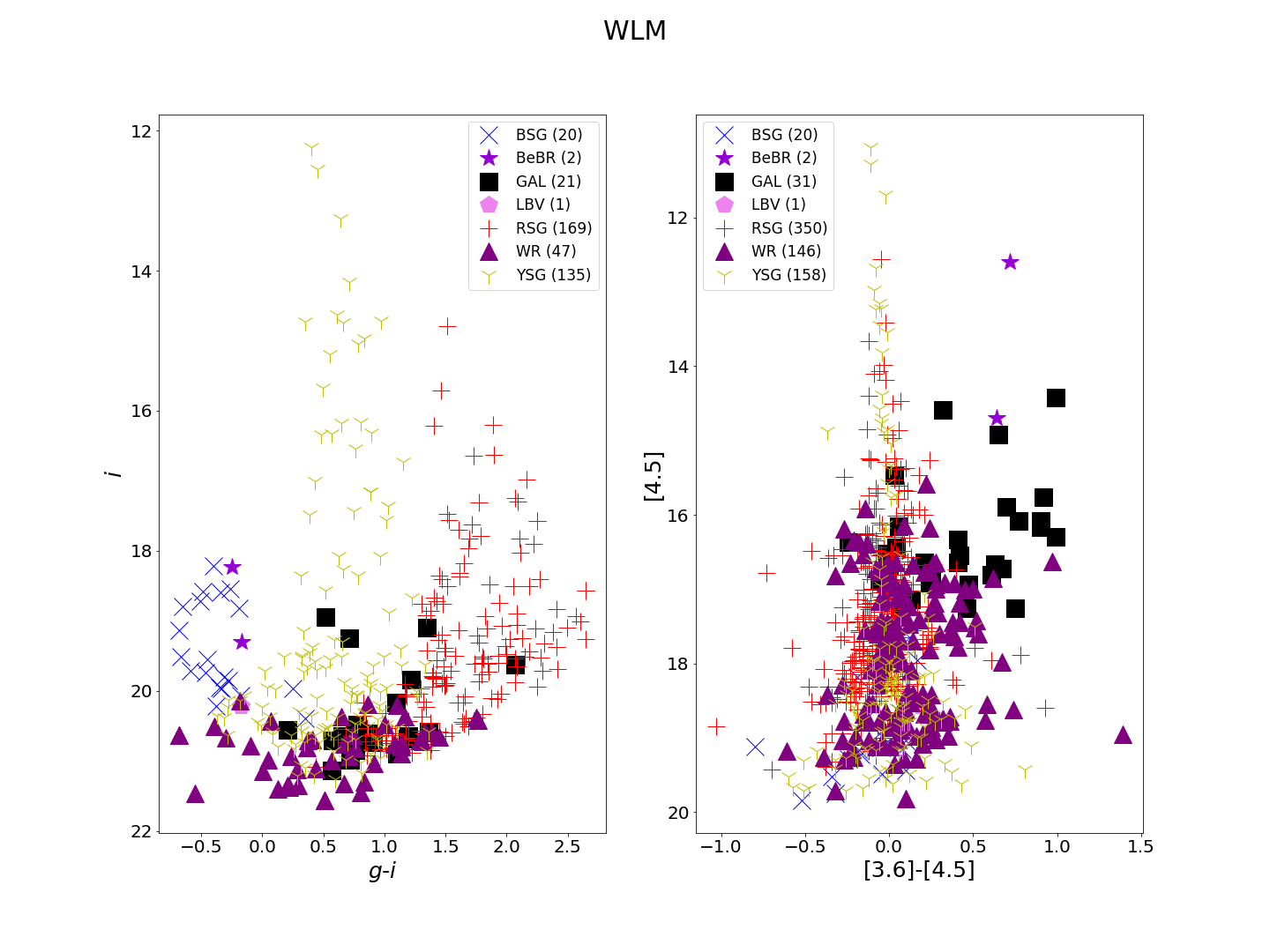} 
 \caption{An optical (left) and IR (right) CMD for WLM. We can easily notice the distributions of BSG (blue 'X'), YSG (yellow slim triangles), RSG (red crosses), WR (purple full triangles), GAL (black boxes), BeBR (dark violet stars), LBV (pink pentagons). Note that the number of sources in the optical CMD is smaller than the IR one (708 sources in total) because not all sources have $g$ magnitudes.}   
  \label{f:cmds}
\end{center}
\end{figure*}

Given the spectral classifications and photometric information we can proceed and plot indicative Color-Magnitude Diagrams (CMDs) to examine our results. In Figure \ref{f:cmds} we plot the optical $g-i$ vs. $i$ and the IR [3.6]-[4.5] vs. [4.5] CMDs for 708 WLM sources. We can easily notice the distributions of sources as expected in the optical with the BSG, YSG, and RSG occupying different distinct locations. There are numerous WR though which seem to overlap with other populations. This is also evident in the IR CMD, and it is a possible effect of a higher rate of false positives for this class. In the IR CMD we sse the clustering of RSG above the BSG populations as expected \citep[c.f.][]{Bonanos2009}. Interestingly, we identify one LBV and two BeBR candidate sources, which shows that the classifier is able to highlight the most promising targets.

\begin{figure}
% \vspace*{-2.0 cm}
\begin{center}
 \includegraphics[width=\textwidth]{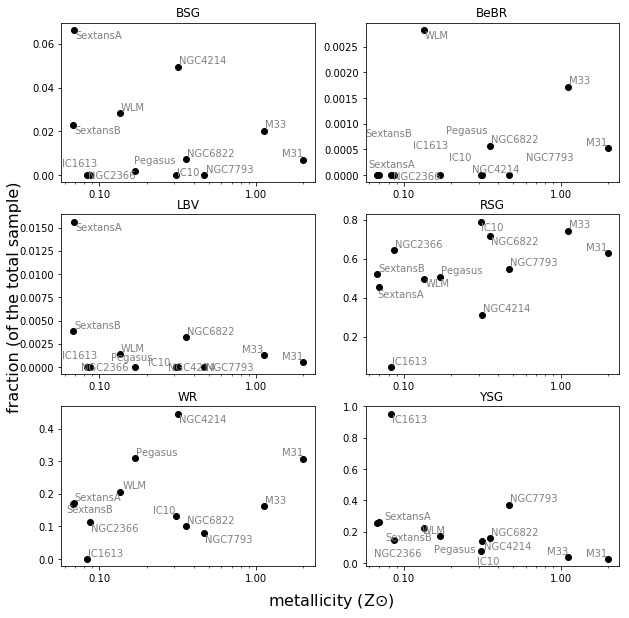}
%\vspace*{-1.0 cm}
 \caption{The fractions, of the predicted class members over the total sample size for each galaxy, with metallicity. }
  \label{f:fractions}
\end{center}
\end{figure}

Having obtained the predictions for all galaxies, we can examine the predictions in a more global framework. We considered the fractions of the number of predicted sources (per class) over the total number of sources in each galaxy. In Figure \ref{f:fractions} we present these fractions with metallicity for a subsample of our galaxies. It is interesting to point out the increasing (decreasing) trend of BSG (RSG) with lower metallicities consistent with theoretical expectations that more (less) time is spend in these phases \citep[e.g.][]{Ren2021, Massey2002}. Also, WR display decreasing trend because of the drop in the strength of the stellar winds with metallicity. YSG display a rather flatter behavior (with the exception of IC 1613 which deserves a revisit), while we notice far fewer BeBR than LBVs in low metallicity galaxies.  

However, we need to be careful with the interpretation of these plots, as currently we have not taken into account the errors with respect to the predictions that should propagate through the end fractions, as well as completeness issues. As our samples are based on \textit{Spitzer} catalogs we are looking for (and predicting) stars with substantial amounts of dust. Although this can be true for the majority of the RSG and the BeBR it is not necessary for BSG (which includes main-sequence massive stars) or YSGs. We also need to quantify the contamination. For example, the BSG class may get mixed with HII regions, while bright AGB stars may still be present among the RSGs. However, at the distances of the galaxies we are looking at we are actually probing the upper part of the Hertzsprung–Russell diagram, which minimizes the contamination by low-mass stars. Last but not least, we need to further study in detail the influences of our selection cuts. For example Fig. 8 in \citep{Maravelias2022} shows that the peak probability of incorrect classification shifts to higher probability values when examining the test galaxies (lower-Z environments). 

\section{Conclusions}

We developed a machine-learning classifier based on optical (Pan-STARRS) and IR (\textit{Spitzer}) photometry, using M31 and M33 sources as training sample, with an overall weighted balanced accuracy of $\sim83\%$. Then we applied this classifier to 25 nearby galaxies (with $1/15 - \sim3~Z_{\odot}$) to get spectral classification (including Blue/Yellow/Red Supergiants, B[e] Supergiants, Wolf-Rayet, Luminous Blue Variables, and background galaxies) for about 1.1 million sources. Selecting some initial quality cuts for the output class probability and data availability we explore the population statistics with respect to the metallicity. We identify potentially interesting trends, such as the increase/decrease of Blue/Red supergiants and drop of WR with decreasing metallicity. However, to fully exploit these results further processing is needed to correctly account for predictions errors, completeness issues, and influences of quality cuts. Nevertheless, this catalog is an excellent tool for identifying interesting objects and prioritizing targets for observations, especially for \textit{JWST}.

%%%%%%%%%%%%%%%%%%%%%%%%%%%%%%%%%%%%%%%%%%%%%%%%%%%%%%%%%%
\section*{Acknowledgments}

We acknowledge funding support from the European Research Council (ERC) under the European Union’s Horizon 2020 research and innovation programme (Grant agreement No. 772086).

%%%%%%%%%%%%%%%%%%%%%%%%%%%%%%%%%%%%%%%%%%%%%%%%%%%%%%%%%%
 \bibliographystyle{apj}
 \bibliography{refs}
%%%%%%%%%%%%%%%%%%%%%%%%%%%%%%%%%%%%%%%%%%%%%%%%%%%%%%%%%%

%%%%%%%%%%%%%%%%%%%%%%%%%%%%%%%%%%%%%%%%%%%%%%%%%%%%%%%%%%
\begin{discussion}

\discuss{Miriam Garcia}{
 I have two questions: a. What about using spectroscopic data in the machine-learning method? and b. In IC 1613 we have detected an LBV-candidate and a WO star, which I do not see in your results.
}
\vspace{-2mm}

\discuss{Grigoris Maravelias}{
 As far as the first question, we want to maximize the number of targets for which we want to have predicted spectral types. The largest lists of stars can come from photometry. In supervised machine learning you need the input data and their corresponding classes for proper training. When you apply the classifier you have to do it in on a dataset which has the same input (in our case photometry in the form of color indices) as your training. That's why we tried to get as many spectral types as possible from literature and link them to photometry. If we would like to use other input data to train (e.g. spectra, spectral parameters) we should have the exact same features to apply to. And unfortunately there is not such a set (and if there was, e.g. spectra, the sample size would be much more limited in contrast to photometry).\\
 For the second question, I actually presented only a part of our results, based on some quality cuts including those without any missing data. However, this would open the discussion about how we handle cases with missing values and I would redirect you to the paper as this demands much more time for discussion. It is most possible that these sources can be found in the other parts.
}
\vspace{-2mm}

\discuss{Griffin Hosseinzadeh}{
 Is it possible to train the algorithm with lower metallicity samples?  
}
\vspace{-2mm}

\discuss{Grigoris Maravelias}{
 With the current use of \textit{Spitzer} and Pan-STARRS photometry only M31 and M33 have significant samples. The Magellanic Clouds are not covered by Pan-STARRS and the inclusion of other surveys would include more systematics. However, this will change as we do have a large observing campaign to obtain spectra in many of these galaxies (along with others of course, e.g. see Marta Lorenzo’s work on Sextans A), so that we can re-train the classifier in the future with more data.
}
\end{discussion}
%%%%%%%%%%%%%%%%%%%%%%%%%%%%%%%%%%%%%%%%%%%%%%%%%%%%%%%%%%

\end{document}